\input harvmac

\noblackbox


\def\IZ{\relax\ifmmode\mathchoice
{\hbox{\cmss Z\kern-.4em Z}}{\hbox{\cmss Z\kern-.4em Z}} {\lower.9pt\hbox{\cmsss Z\kern-.4em Z}}
{\lower1.2pt\hbox{\cmsss Z\kern-.4em Z}}\else{\cmss Z\kern-.4em Z}\fi}
\def\IB{\relax{\rm I\kern-.18em B}}
\def\IC{{\relax\hbox{\kern.3em{\cmss I}$\kern-.4em{\rm C}$}}}
\def\ID{\relax{\rm I\kern-.18em D}}
\def\IE{\relax{\rm I\kern-.18em E}}
\def\IF{\relax{\rm I\kern-.18em F}}
\def\IG{\relax\hbox{$\inbar\kern-.3em{\rm G}$}}
\def\IGa{\relax\hbox{${\rm I}\kern-.18em\Gamma$}}
\def\IH{\relax{\rm I\kern-.18em H}}
\def\II{\relax{\rm I\kern-.18em I}}
\def\IK{\relax{\rm I\kern-.18em K}}
\def\IP{\relax{\rm I\kern-.18em P}}

\font\cmss=cmss10 \font\cmsss=cmss10 at 7pt
\def\IR{\relax{\rm I\kern-.18em R}}

\def\frac#1#2{{#1 \over #2}}

\def\OL#1{ \kern1pt\overline{\kern-1pt#1
   \kern-1pt}\kern1pt }

\lref\adsorb{
S.~Kachru and E.~Silverstein, ``4d conformal theories and strings on orbifolds,'' Phys.\ Rev.\ Lett.\  {\bf 80},
4855 (1998) [arXiv:hep-th/9802183].
}

\lref\quiver{
M.~R.~Douglas and G.~W.~Moore, ``D-branes, Quivers, and ALE Instantons,'' arXiv:hep-th/9603167.
}

\lref\tachconnect{
S.~P.~de Alwis, J.~Polchinski and R.~Schimmrigk, ``Heterotic Strings With Tree Level Cosmological Constant,''
Phys.\ Lett.\ B {\bf 218}, 449 (1989);
R.~C.~Myers, ``New Dimensions For Old Strings,'' Phys.\ Lett.\ B {\bf 199}, 371 (1987);
A.~H.~Chamseddine, ``A Study of noncritical strings in arbitrary dimensions,'' Nucl.\ Phys.\ B {\bf 368}, 98
(1992);
N.~Berkovits, A.~Sen and B.~Zwiebach, ``Tachyon condensation in superstring field theory,'' Nucl.\ Phys.\ B {\bf
587}, 147 (2000) [arXiv:hep-th/0002211];
S.~Kachru, J.~Kumar and E.~Silverstein, ``Orientifolds, RG flows, and closed string tachyons,'' Class.\ Quant.\
Grav.\  {\bf 17}, 1139 (2000) [arXiv:hep-th/9907038];
 S.~Hellerman, ``On the landscape of superstring theory in D > 10,'' arXiv:hep-th/0405041.
}

\lref\FengIF{ J.~L.~Feng, J.~March-Russell, S.~Sethi and F.~Wilczek, ``Saltatory relaxation of the cosmological
constant,'' Nucl.\ Phys.\ B {\bf 602}, 307 (2001) [arXiv:hep-th/0005276].
}

\lref\GKP{S. Giddings, S. Kachru, and J. Polchinski, ``Hierarchies from Fluxes in String Compactifications,''
Phys. Rev. {\bf D66} (2002) 106006, hep-th/0105097. }

\lref\dinewitt{
M.~Dine, P.~J.~Fox and E.~Gorbatov, ``Catastrophic decays of compactified space-times,'' arXiv:hep-th/0405190.
}

\lref\bankszaks{
T.~Banks and A.~Zaks, ``On The Phase Structure Of Vector - Like Gauge Theories With Massless Fermions,'' Nucl.\
Phys.\ B {\bf 196}, 189 (1982).
}

\lref\susyscale{
T.~Banks, M.~Dine and E.~Gorbatov, ``Is there a string theory landscape?,'' arXiv:hep-th/0309170;
M.~Dine, E.~Gorbatov and S.~Thomas,
arXiv:hep-th/0407043;
M.~R.~Douglas, ``Statistical analysis of the supersymmetry breaking scale,'' arXiv:hep-th/0405279.
L.~Susskind, ``Supersymmetry breaking in the anthropic landscape,'' arXiv:hep-th/0405189;
N. Arkani-Hamed, S. Dimopoulos, and S. Kachru, work in progress.}

\lref\splitsusy{
N.~Arkani-Hamed and S.~Dimopoulos, ``Supersymmetric unification without low energy supersymmetry and signatures
for fine-tuning at the LHC,'' arXiv:hep-th/0405159.
}

\lref\perlmutter{
S.~Perlmutter {\it et al.}  [Supernova Cosmology Project Collaboration], ``Measurements of Omega and Lambda from
42 High-Redshift Supernovae,'' Astrophys.\ J.\  {\bf 517}, 565 (1999) [arXiv:astro-ph/9812133].
}

\lref\other{S. Gukov, C. Vafa and E. Witten, ``CFTs from Calabi-Yau Fourfolds,'' Nucl. Phys. {\bf B584} (2000)
69, hep-th/9906070\semi T.R. Taylor and C. Vafa, ``RR flux on Calabi-Yau and partial supersymmetry breaking,''
Phys. Lett. {\bf B474} (2000) 130, hep-th/9912152\semi P. Mayr, ``On Supersymmetry Breaking in String Theory and
its Realization in Brane Worlds,'' Nucl. Phys. {\bf B593} (2001) 99, hep-th/0003198; A. Strominger,
``Superstrings with Torsion,'' Nucl. Phys. {\bf B274} (1986) 253\semi J. Polchinski and A. Strominger, ``New
vacua for type II string theory,'' Phys. Lett. {\bf B388} (1996) 736, hep-th/9510227\semi K. Becker and M.
Becker, ``M-theory on eight manifolds,'' Nucl. Phys. {\bf B477} (1996) 155, hep-th/9605053\semi
S. Kachru,~M. Schulz and S. P. Trivedi, ``Moduli Stabilization from Fluxes in a Simple IIB Orientifold,'' JHEP
{\bf 0310} (2003) 007, hep-th/0201028\semi A. Frey and J. Polchinski, ``N=3 Warped Compactifications,'' Phys.
Rev. {\bf D65} (2002) 126009, hep-th/0201029; P. Tripathy and S. P. Trivedi, ``Compactification with Flux on K3
and Tori,''  JHEP {\bf 0303} (2003) 028, hep-th/0301139; A. Giryavets, S. Kachru, P. Tripathy and S. Trivedi,
``Flux Compactifications on Calabi-Yau Threefolds,'' JHEP {\bf 0404} (2004) 003, hep-th/0312104;
J.Michelson, ``Compactifications of type IIB strings to four-dimensions with non-trivial classical potential,''
Nucl. Phys. {\bf B495} (1997) 127, hep-th/9610151\semi K. Dasgupta, G. Rajesh and S. Sethi, ``M-theory,
orientifolds and G-flux,'' JHEP {\bf 9908} (1999) 023, hep-th/9908088\semi B. Greene, K. Schalm and G. Shiu,
``Warped compactifications in M and F theory,'' Nucl. Phys. {\bf B584} (2000) 480, hep-th/0004103\semi G. Curio,
A. Klemm, D. L\"ust and S. Theisen, ``On the Vacuum Structure of Type II String Compactifications on Calabi-Yau
Spaces with H-Fluxes,'' Nucl. Phys. {\bf B609} (2001) 3, hep-th/0012213\semi K. Becker and M. Becker,
``Supersymmetry Breaking, M Theory and Fluxes,'' JHEP {\bf 010} (2001) 038, hep-th/0107044\semi M. Haack and J.
Louis, ``M theory compactified on Calabi-Yau fourfolds with background flux,'' Phys. Lett. {\bf B507} (2001)
296, hep-th/0103068\semi J. Louis and A. Micu, ``Type II theories compactified on Calabi-Yau threefolds in the
presence of background fluxes,'' Nucl.Phys. {\bf B635} (2002) 395, hep-th/0202168. M. Becker, G. Curio and A.
Krause, ``De Sitter Vacua from Heterotic M Theory,'' hep-th/0403027\semi R. Brustein and S.P. de Alwis, ``Moduli
Potentials in String Compactifications with Fluxes: Mapping the Discretuum,'' hep-th/0402088\semi S. Gukov, S.
Kachru, X. Liu and L. McAllister, ``Heterotic Moduli Stabilization with Fractional Chern-Simons Invariants,''
hep-th/0310159\semi E. Buchbinder and B. Ovrut, ``Vacuum Stability in Heterotic M-theory,'' hep-th/0310112.}

\lref\atish{
A.~Dabholkar, A.~Iqubal and J.~Raeymaekers, ``Off-shell interactions for closed-string tachyons,'' JHEP {\bf
0405}, 051 (2004) [arXiv:hep-th/0403238].
}

\lref\twistedcirc{
J.~R.~David, M.~Gutperle, M.~Headrick and S.~Minwalla, ``Closed string tachyon condensation on twisted
circles,'' JHEP {\bf 0202}, 041 (2002) [arXiv:hep-th/0111212].
}

\lref\barton{
Y.~Okawa and B.~Zwiebach, ``Twisted tachyon condensation in closed string field theory,'' JHEP {\bf 0403}, 056
(2004) [arXiv:hep-th/0403051].
}

\lref\BP{
R.~Bousso and J.~Polchinski, ``Quantization of four-form fluxes and dynamical neutralization of the cosmological
constant,'' JHEP {\bf 0006}, 006 (2000) [arXiv:hep-th/0004134].
}

\lref\klebtach{
I.~R.~Klebanov and A.~A.~Tseytlin, ``D-branes and dual gauge theories in type 0 strings,'' Nucl.\ Phys.\ B {\bf
546}, 155 (1999) [arXiv:hep-th/9811035].
}

\lref\aps{
A.~Adams, J.~Polchinski and E.~Silverstein, ``Don't panic! Closed string tachyons in ALE space-times,'' JHEP
{\bf 0110}, 029 (2001) [arXiv:hep-th/0108075]; C.~Vafa, ``Mirror symmetry and closed string tachyon
condensation,'' arXiv:hep-th/0111051;
J.~A.~Harvey, D.~Kutasov, E.~J.~Martinec and G.~Moore, ``Localized tachyons and RG flows,''
arXiv:hep-th/0111154;
see M.~Headrick, S.~Minwalla and T.~Takayanagi, ``Closed string tachyon condensation: An overview,'' Class.\
Quant.\ Grav.\  {\bf 21}, S1539 (2004) [arXiv:hep-th/0405064] for a nice review with many other useful
references.

}

\lref\adstach{
A.~Adams and E.~Silverstein, ``Closed string tachyons, AdS/CFT, and large N QCD,'' Phys.\ Rev.\ D {\bf 64},
086001 (2001) [arXiv:hep-th/0103220];
A.~A.~Tseytlin and K.~Zarembo, ``Effective potential in non-supersymmetric SU(N) x SU(N) gauge theory and
interactions of type 0 D3-branes,'' Phys.\ Lett.\ B {\bf 457}, 77 (1999) [arXiv:hep-th/9902095];
D.~Tong, ``Comments on condensates in non-supersymmetric orbifold field theories,'' JHEP {\bf 0303}, 022 (2003)
[arXiv:hep-th/0212235].
}

\lref\distributions{S.~Ashok and M.~R.~Douglas, ``Counting flux vacua,'' JHEP {\bf 0401}, 060 (2004)
[arXiv:hep-th/0307049]. \
F.~Denef and M.~R.~Douglas,
JHEP {\bf 0405}, 072 (2004) [arXiv:hep-th/0404116]. \
A.~Giryavets, S.~Kachru and P.~K.~Tripathy, ``On the taxonomy of flux vacua,'' arXiv:hep-th/0404243.}

\lref\kklt{ S.~Kachru, R.~Kallosh, A.~Linde and S.~P.~Trivedi, ``De Sitter vacua in string theory,''
arXiv:hep-th/0301240.
}

\lref\mss{ E.~Silverstein, ``(A)dS backgrounds from asymmetric orientifolds,'' arXiv:hep-th/0106209;
A.~Maloney, E.~Silverstein and A.~Strominger, ``De Sitter space in noncritical string theory,''
arXiv:hep-th/0205316.
}

\input epsf
\noblackbox
\newcount\figno
\figno=0
\def\fig#1#2#3{
\par\begingroup\parindent=0pt\leftskip=1cm\rightskip=1cm\parindent=0pt
\baselineskip=11pt \global\advance\figno by 1 \midinsert \epsfxsize=#3 \centerline{\epsfbox{#2}} \vskip 12pt
{\bf Fig.\ \the\figno: } #1\par
\endinsert\endgroup\par
}
\def\figlabel#1{\xdef#1{\the\figno}}

\Title{\vbox{\baselineskip12pt\hbox{hep-th/0407202} \hbox{SLAC-PUB-10569}\hbox{SU-ITP-04/30} }}
{{\centerline{Counter-intuition and Scalar Masses}}}

\centerline{Eva Silverstein}
\bigskip

\centerline{SLAC and Department of Physics, Stanford University, Stanford, CA 94305/94309}

\vskip .3in \centerline{\bf Abstract} { The Bousso-Polchinski mechanism for discretely fine-tuning the
cosmological constant favors a large bare negative cosmological constant.  I argue (using generalizations of
results of Klebanov and Tseytlin) that a similar mechanism for fine-tuning the scalar masses to small values
favors a large bare negative (i.e. tachyonic) scalar mass.  I comment briefly on the related issue of the role
of low energy supersymmetry in string theory.

} \vskip .1in

\smallskip
\Date{July 2004}


\vfill \eject

\newsec{The c.c. and $m_T$}

In \BP, Bousso and Polchinski proposed that in the presence of multiple flux quantum numbers arising from form
fields on compactification geometries, the cosmological constant can be finely tuned to a value near zero (see
also \FengIF). They modelled this mechanism in a very simple way, via a cancellation between a bare negative
cosmological constant and positive contributions from form field kinetic terms integrated over a
compactification:
\eqn\BPcanc{\Lambda_{BP}=-|\lambda_{bare}|+\sum_i c_i Q_i^2.}
Here $i$ labels cycles on the compactification geometry, which carry flux quantum numbers $Q_i$ and contribute
to the cosmological term via the form field kinetic terms. $\lambda_{bare}$ contains the rest of the
contributions to the cosmological constant, and the mechanism applies when this is negative. In \BP, the
coefficients $c_i$ (with incommensurate values of $\sqrt{c_i}$) were taken to be constants sufficiently smaller
than one to have control over the $\alpha'$ expansion.\foot{though in geometrical compactifications realizing
the idea in \refs{\GKP,\kklt}\ these coefficients are themselves nonlinear functions of the fluxes.}. As
explained in \BP, a lattice of fluxes of dimensionality $b\ll Q$ yields of order $Q^{b-1}$ vacua in a shell of
radius $Q$ and unit thickness in flux space. As a result, with a multiplicity of fluxes one can discretely tune
the total cosmological constant $\Lambda_{BP}$ to a small value of order $Q^{-b}$ in Planck units. It is then
evident that within the framework of this simple model \BPcanc, there are more vacua with small cosmological
constant if one starts with a larger magnitude of $\lambda_{bare}$, since the radius of the corresponding shell
in flux space is larger.

In this note, I will point out a parallel argument for the distribution of values of scalar masses.  After
summarizing this in the next subsection, I will move on to discuss caveats to the simple picture \BPcanc\ in
\S1.2.  Finally in \S2\ I will explain a slightly more elaborate setup in which the mechanism for tuning scalar
masses can be combined with the Higgs mechanism.  Along the way I will describe the generalization of stringy
tachyon couplings \klebtach\ to more general examples \aps.

\subsec{Scalar Masses}

Now let us move from the cosmological constant to scalar masses. In \klebtach, it was argued in the context of
the AdS/CFT correspondence that RR fluxes couple to closed string tachyons (specifically the twisted tachyon $T$
of type II orbifolded by spacetime fermion number, otherwise known as ``type 0") via the interaction
\eqn\KTach{{\cal L}_{TF}\sim -|T|^2(\sum_I |F_I|^2),}
where $I$ indexes different types of RR flux.  This goes in the direction of stabilizing the tachyon (decreasing
the magnitude of its negative mass squared in a nontrivial flux background).  In \S2\ I will argue that a
similar effect exists for more general twisted tachyons \aps.\foot{In the AdS/CFT context, one can analyze the
vacuum structure at weak coupling to see if this effect actually stabilizes $T$ in the limit of small radius and
large flux density. One finds \adstach\ that the instability has become milder at weak field theory coupling
(though a corresponding instability evident in the one loop Coleman Weinberg potential remains).}

The combination of the bare mass and flux contribution, applied to a compactification $X$, yields a mass formula
of the form
\eqn\tachmass{m_T^2 = -\eta m_s^2 + \sum_I \tilde c_I{|Q_I|^2}+\dots \equiv m_{bare}^2+\sum_I \tilde
c_I{|Q_I|^2}}
where the $Q_I$ represent the independent flux quantum numbers (indexed by $I$) of the RR fields coupling to the
tachyon as in \KTach, and as in \BPcanc\ the moduli-dependent coefficients $\tilde c_I$ arise from the
dimensional reduction on $X$. Here $m_s$ is the string mass scale, and the term $-\eta m_s^2$ represents the
tree level tachyon mass squared (which in perturbative string theory ranges down to $\eta$ of order one); the
$\dots$ represent other contributions to the tachyon mass which may arise depending on the background
considered.  We have included these other contributions with the $-\eta m_s^2$ into a ``bare" mass term; we will
be interested in regions in the landscape in which $m_{bare}^2<0$ so that the positive flux contribution can
tune the total $m_T^2$ close to zero.

This formula \tachmass\ is of the same form as the formula \BPcanc\ for the structure of the cosmological term
\BP.  In particular, if one wishes to tune away the scalar mass to a small value, it is advantageous to start
with a scalar such as $T$ with a large bare negative mass squared.  That is, as in the original
Bousso-Polchinski argument \BPcanc, there are more choices of flux which tune away a larger bare negative value
than ones which tune away a smaller bare negative value.

Tuning a scalar mass to a small value can also have the effect of tuning its VEV to a small value.  For example,
consider the cases with a stabilizing positive quartic potential term $g_4 |T|^4$ for $T$.  Independently tuning
the quadratic term \tachmass\ to a small negative value $|m_T^2|\ll m_s^2 g_4$ produces a VEV of order
\eqn\TVEV{\langle T\rangle \sim |m_T|/\sqrt{g_4}\ll m_s}

It would be interesting to apply this observation to scalar fields arising in models of the real world.  One
application might be to cosmological models requiring light scalar fields.  The other obvious target for this
mechanism is the Higgs particle of the standard model. In order to apply this observation to the Higgs, one
requires a more realistic scalar field than the type 0 tachyon. Therefore in \S2, I will turn to a slightly more
elaborate setting in which this mechanism applies to gauge-charged scalar fields.

\subsec{Further developments and important caveats}

Before proceeding to that, let me briefly review some further aspects of this line of development, including
important qualifications applying to the simple models \BPcanc\tachmass\ considered above.

The Bousso-Polchinski mechanism, while a compelling proposal especially given the presence of the basic
ingredients (flux and compactification geometry) in string theory, suffered at the time from the lack of
examples exhibiting sufficient compensating forces to plausibly fix all the moduli.  That problem has been
addressed in various special limits of string theory, specifically supercritical type II limits \mss\ and
Calabi-Yau flux compactifications of the critical type IIB limit \refs{\GKP,\kklt}\ (see also \other\ for
important work toward fixing all the moduli in various limits of the theory).

These models realize the spirit of the Bousso Polchinski mechanism, albeit with some important differences from
the simple model \BPcanc.  In the supercritical examples of \mss, an explicit asymmetric orientifold fixes at
tree level all of the runaway moduli except the dilaton. The dilaton is fixed by the contributions to the
dilaton potential from supercriticality \tachconnect, orientifolds, and RR fluxes.  This potential in the
specific asymmetric orientifold models of \mss\ tunes the cosmological constant to small values by using a large
$D$ expansion rather than directly the Bousso Polchinski flux lattice, since in these particular models the flux
lattice is too regular to provide by itself a finely spaced discretuum.  In the examples of \kklt, the
moduli-dependent coefficients $c_i$ are themselves complicated functions of the fluxes since the fluxes help
determine where the moduli get stabilized.  As a result, the form \BPcanc\ is at best schematic. Similarly, in
\tachmass\ in general the coefficients $\tilde c_I$ will be nonlinearly related to the fluxes.  However, as
emphasized in the recent works \susyscale, the issue is not so much the precise form of the positive terms in
\BPcanc\ (and now also \tachmass) but their independent distribution.

Another important caveat to simple arguments such as those reviewed above, as explained in \susyscale, is that
there may also be branches of solutions with vanishing coefficients of the classical contributions in
\BPcanc\tachmass, such as cases in which the cosmological term and scalar masses are completely generated by
non-perturbative effects. The question then becomes whether these cases are more numerous than those considered
above.  Similarly, in the case of the scalar masses in general we must consider off diagonal elements in the
mass matrix, and their couplings to fluxes, which may yield a different accounting of preferred bare
values.\foot{I thank S. Dimopoulos for discussions on this issue.}

Finally, having stated these caveats, it is worth emphasizing that the Bousso-Polchinski argument naturally
predicts a {\it nonzero} value for the cosmological constant among vacua without exact extended supersymmetry,
since the discretuum produced by the quantized flux quantum numbers leads to the positive terms in \BPcanc\
being discretely but not continuously tunable.  As a result, the data \perlmutter\ fits in well qualitatively
with their mechanism in general and string theory in particular.

\newsec{Applications}

One application for the scalar mass analogue of Bousso-Polchinski is to models of inflation which require a
light scalar field.  Another is to the Higgs particle in the Standard Model.  For both these applications,
particularly the latter, it is of interest to generalize the mechanism in \S1.2\ beyond the Type 0 setting in
which the couplings \KTach\ were motivated \klebtach.

\subsec{More general tachyons and their RR couplings}

A set of generalizations of the Type 0 tachyon is twisted tachyons in $Z_k$ orbifolds of flat space \aps. These
are closely related to the type 0 tachyon as can be seen from the large $k$ limit \refs{\aps,\atish}\ and the
``twisted circle" examples where the $Z_k$ rotation is accompanied by a shift on another dimension \twistedcirc.
They also arise in nonsupersymmetric orbifolds of $AdS/CFT$ \refs{\adsorb,\adstach}, where as in
\refs{\klebtach,\adstach}\ one finds that the small radius large flux regime has a milder instability than that
arising at large radius. As a result, we expect that these localized tachyons also have couplings to RR fluxes
of the form \KTach. This could be checked via an analogous worldsheet computation to \KTach\klebtach, with the
same ambiguities regarding the off shell nature of the zero momentum tachyon; these couplings are not prohibited
by any selection rules from the worldsheet symmetries. It could in principle be checked off shell in string
field theory, along the lines of \barton, though the superstring case is still somewhat out of reach.

In fact there is a simpler argument that the coupling \KTach\ persists to the case of tachyons localized by
geometrical orbifolds, at least in a regime accessible to a general relativistic analysis (which occurs far
enough along the flow in \aps\ and can be enforced also by resolving the tip of the cone independently with an
extra shift \twistedcirc). This follows from considering the basic form of the spacetime deformation induced by
the localized tachyon condensation \aps, for which we will then consider the effects of adding RR flux. The
tachyon condensation produces a smoothing of the tip of the cone inside a shell of dilaton/graviton (outside of
which is the remaining part of the original solution); see figure 1 of \aps. This deformation has the property
that the region inside the shell is of smaller volume after the tachyon condensation than the corresponding
region before the condensation (see the discussion below equation (4.9) in \aps).

In order to understand the analogue of the coupling \KTach\ in the localized case, we then need to determine the
effect of bulk RR flux on this expanding shell configuration.  For realistic application, this tachyonic cone
must be embedded in a compact manifold, and the RR fluxes integrated over cycles in this compactification
manifold are quantized. Since the integral of a bulk RR flux over the cycles it threads is quantized, the
integral of this flux over the region of space inside the shell is constant (since the region outside the shell
is identical to the original orbifold background). Since, as we just reviewed, the tachyon condensation lowers
the volume of space inside the shell, the energy density contained in the flux increases upon tachyon
condensation.  Thus the flux pushes the tachyon back toward $T=0$.  Because we are considering bulk RR fluxes,
which come from the untwisted sector of the orbifold, the quantum symmetry of the orbifold prohibits any linear
coupling of $T$ to these fluxes, but permits a coupling of the form \KTach.

I should emphasize that this coupling of the form \KTach\ in the case of the localized tachyons will only occur
for those fluxes which have support where the tachyon is localized.  This reduces the number of independent
fluxes contributing to \KTach\ relative to the Type 0 case, but can still permit large numbers of such fluxes.

\subsec{Bousso-Polchinski for charged scalars}

In order to apply this tuning to the Higgs mechanism, we require a scalar field charged under a non-abelian
gauge group.  This is not the case even for the more general tachyons \aps\ near the orbifold point, but these
tachyons can implement this mechanism indirectly through their couplings to charged scalar fields on D-branes.

In the presence of D-branes at the (blown up) orbifold, the tachyon VEVs $T$ determine the VEVs of D-brane
scalar fields $\phi$. That is, at leading order in $T$ (where $T=0$) is the orbifold limit) they contribute
couplings of the form $-|T|\phi^2$ for D-brane probes \aps.  Hence a small value for the VEV of $T$ leads to a
small scale of symmetry breaking for the D-brane Higgs fields $\phi$.

For example, D-brane probes of $\IC/\IZ_k$ leads to a quiver theory \quiver\ including a gauge group $U(N)^k$
with bifundamental scalar matter $Z_{j,j+1}$ with $j=1,\dots, k$ indexing the factors in the gauge group.
Including the leading effects of the twisted tachyons which in a convenient basis we will label $T_j$
(satisfying $\sum_j T_j=0$), one obtains the following potential for the charged scalars $Z$ in the presence of
a fixed background $T_j$ \aps\
\eqn\Zpot{V(Z)=\sum_j\left( |Z_{j,j+1}|^2-|Z_{j-1,j}|^2+T_j \right)^2+{\cal O}(T^2)}
This leads generically to a Higgs mechanism breaking $U(N)^k$ to $U(N)$ (with special subspaces leading to
$U(N)^{k-2}$, $U(N)^{k-4}$, etc.).  The value of the tachyon VEVF determined by \tachmass\TVEV\ thus also
governs the VEVs of the brane Higgs fields $Z$.

Thus, in simple situations with stringy tachyons, the tachyon VEV controls the VEVs of D-brane Higgs fields.
Since the number of flux choices leading to a small VEV \TVEV\ for $T$ is greater for greater bare negative
tachyon mass squared, this generalization of the Bousso Polchinski mechanism to Higgs scalar masses favors
string backgrounds with a highly tachyonic bare mass term.

\newsec{Discussion}

It is interesting that the mechanism explored in this paper in itself favors backgrounds which include those
closely related to tachyonic non-supersymmetric perturbative string theories -- starting points which are
ordinarily discarded in considering application to the real world.  Of course in order to reach definite
conclusions about statistically favored vacua of string theory, one needs much more comprehensive information
about the full space of vacua, as well as ultimately dynamical information affecting vacuum selection.

There have been several very interesting forays recently into the statistics of supersymmetry breaking in
various corners of M theory \susyscale, related to studies of distributions of IIB flux vacua \distributions\
and to new phenomenological ideas involving high scale supersymmetry breaking \splitsusy. The studies
\susyscale\ aim in part at determining whether string theory in itself (without phenomenological input) can be
seen to predict in a statistical sense either low or high scale supersymmetry breaking.  In other words, they
aim to determine whether more tuning of the UV string theory is required to enforce low energy supersymmetry
(having tuned the cosmological constant) than is required to tune the cosmological constant without low energy
supersymmetry.

In this regard, it is important to note that there are many starting points with string scale or higher
supersymmetry breaking, which were neglected in most of \susyscale, whose contribution to the statistics must be
included before a reliable accounting of the generic scale of supersymmetry breaking can be made. For example,
the addition of spacefilling branes and antibranes, and non-critical dimensionality (dimension $D\ne 10$ for the
superstring), arise as consistent possibilities in perturbative string theory, with some such examples related
by tachyon condensation to the better studied critical string backgrounds with low energy supersymmetry
\tachconnect. These a priori more generic starting points admit sufficient forces to fix their moduli and avoid
tachyons in perturbation theory at weak bare string coupling \mss.\foot{They may generically be subject to
non-perturbative decays, as in \kklt, or in some special cases \dinewitt, but this would still allow them to
persist for timescales exponentially long as a function of appropriate coupling constants.} These intrinsically
string-scale SUSY breaking directions in discrete parameter space threaten to dwarf the statistical contribution
of the special case of critical low energy supersymmetric models; for example the number of flux choices grows
exponentially with dimensionality $D$, as does the number of independent topological quantum numbers of
compactification geometries.\foot{More simply: the statistics of string theory papers may be quite different
from the statistics of string theory vacua.  For example, ten and two dimensional target spaces have been by far
the most popular, but the other dimensions are available.}

One argument often made for supersymmetry is its enhancement of theoretical control; indeed it has led to
tour-de-force calculations of the two derivative effective action of some interesting theories. On the other
hand, in the context of the problem of fixing the moduli it can happen that non-renormalization theorems simply
postpone needed moduli-fixing contributions to highly subleading (e.g. non-perturbative) orders in an expansion
about weak coupling.  In fixing moduli, ultimately one balances different orders in this expansion off each
other, and in some ways it is simpler to implement this in situations where the needed forces arise at early
orders in perturbation theory (as in \mss). As in the analogous perturbative analysis of gauge theories
\bankszaks, a perturbative balancing of contributions at different orders can produce well controlled
non-supersymmetric backgrounds via the introduction of large discrete quantum numbers.  In any case, theoretical
control does not in itself provide an argument for an observational prediction.

Despite these difficulties, I find the statistical program \refs{\BP,\distributions,\susyscale}\ for seeking
generic properties of string vacua extremely interesting, particularly in its prospects for refining our notions
of naturalness.  The mechanism discussed in this note provides a new twist on these issues in the regions of the
landscape to which it applies.

\noindent{\bf Acknowledgements}

I would like to thank S. Dimopoulos, M. Dine, M. Douglas, W. Fischler, S. Kachru, A. Lawrence, L. Susskind, and
S. Thomas for useful and enjoyable discussions in this area. I would also like to thank the organizers and
participants in the Banff, Johns Hopkins, and CERN high energy theory workshops.  Financial support comes in
part from the DOE under contract DE-AC03-76SF00515 and by the NSF under contract 9870115.

\listrefs

\end